\documentclass[showpacs,showkeys,prb,twocolumn]{revtex4}
\usepackage{amssymb}
\usepackage{amsfonts}
\usepackage{amsmath}
\usepackage{epsfig}
\usepackage{graphicx}

\providecommand{\U}[1]{\protect\rule{.1in}{.1in}}

\begin{document}

\title[Unbalanced Fermi gas]{Superfluidity of a spin-imbalanced Fermi gas in
a three-dimensional optical lattice}
\author{R. Mendoza}
\affiliation{Posgrado en Ciencias F\'{\i}sicas, UNAM; Instituto de F\'{\i}sica, UNAM}
\author{Mauricio Fortes}
\affiliation{Instituto de F\'{\i}sica, UNAM, Apdo. Postal 20-364, 01000 M\'exico D.F.,
M\'exico}
\author{Zlatko Koinov}
\affiliation{Department of Physics and Astronomy, University of Texas at San Antonio, San
Antonio, Texas 78249, USA}
\author{M. A. Sol\'{\i}s}
\affiliation{Instituto de F\'{\i}sica, UNAM, Apdo. Postal 20-364, 01000 M\'exico D.F.,
M\'exico}

\begin{abstract}
We study fermion pairing in a population-imbalanced mixture of $^{6}$Li
atomic gas loaded in a three-dimensional lattice at very low temperatures.
Using the number equation for each population, the gap equation and the
equation for the Helmholtz free energy, we determine the gap, chemical
potentials and pair-momentum as functions of polarization. These parameters
define the stability regions for: a Fulde-Ferrell-Larkin-Ovchinnikov phase; a
phase separation region where BCS and normal phases coexist; a Sarma phase
when the pair-momentum vanishes, and the transition to the normal phase when
the gap disappears. The collective-mode energies are then calculated using a
Bethe-Salpeter approach in the generalized random phase approximation assuming
that the system is well described by the single-band Hubbard model. A novel
result is that this fermionic gas has a superfluid phase revealed by rotonlike minima in the asymmetric collective-mode energy spectrum.
\end{abstract}

\maketitle



\section{Introduction}

The ability to use optical lattices to study the properties of ultracold
atoms provides a testing model to simulate different strongly-correlated
Fermi systems. Optical lattices are also tailored-made to study the effects
of dimensionality on correlated Fermi systems as the former are created by
standing laser waves in one, two or three dimensions \cite{Greiner}. Since
the frequency and intensity of the lasers can be tuned up to specific
values, the properties of ultracold Fermi or Bose systems loaded onto these
lattices can be studied with impressive detail \cite{Esslinger}. In
addition, when the atoms are near a Feshbach resonance their interaction can
be finely tuned to explore the crossover from the weakly-interacting
Bardeen-Cooper-Schrieffer (BCS) regime characterized by Cooper pair
formation to the strongly-interacting regime where the formation of
molecular pairs with zero spin can undergo a Bose-Einstein condensation
(BEC) at sufficiently low temperatures \cite{Revfesch}. Although most
experimental and theoretical models of correlated Fermi systems have dealt
with balanced populations of spin states, more recently \cite{Ketterle2, Liao} the ability to manipulate ultracold atomic clouds has motivated the interest to study systems when the mixture of two
hyperfine states in, for example, an atomic Fermi gas is not balanced. In this case, the
two Fermi surfaces are no longer aligned and the lowest energy pairs have
non-zero total momenta. Such phases were first studied by Fulde and Ferrell
(FF) \cite{FF}, who used an order parameter that varies as a single plane
wave, and by Larkin and Ovchinnikov (LO) \cite{LO}, who suggested that the
order parameter is a superposition of two plane waves.

Although the FF and LO phases (presently referred as FFLO) were introduced quite a long time ago, they
are still of very high interest because the question whether the
superconductivity/superfluidity can survive in 3D polarized systems remains
experimentally unanswered. In the FFLO phase, Cooper pairing occurs between
a fermion with momentum $\boldsymbol{k+q}$ and spin $\uparrow $ and a
fermion with momentum $\boldsymbol{-k+q}$, and spin $\downarrow $. As a
result, the total pair momentum is $2\boldsymbol{q}$ and the order parameter
becomes spatially dependent as proposed by Larkin and Ovchinnikov \cite{LO}.
The mean-field treatment of the FFLO phase in a variety of systems, such as
superconductors with Zeeman splitting and heavy-fermion superconductors \cite%
{SC}, atomic Fermi gases with population imbalance loaded in optical
lattices \cite{Kop1,Kop,FG} and harmonic traps \cite{HT}, and dense quark
matter \cite{DM} shows that the FFLO phase competes with a number of other
phases, such as the Sarma ($\boldsymbol{q}=0$) states \cite{S}, but in some
regions of momentum space the FFLO phase is more stable as it provides the
minimum of the mean-field expression of the Helmholtz free energy.

In this paper we calculate the polarization dependence of the gap, chemical
potentials and pair-momentum as well as the collective excitations of an imbalanced mixture of two hyperfine states $|\!\!\uparrow >$ and $|\!\!\downarrow >$ of a $^{6}$Li atomic
Fermi gas 
under an attractive contact interaction loaded into a cubic optical
lattice.

In Section II we summarize the properties of the Hubbard Hamiltonian used
here to model a two-component Fermi gas in a lattice
produced by standing waves of three pairs of counter-propagating laser beams.
Section III is devoted to the calculation of the thermodynamic potential of
the system. When the number of particles is fixed, the Helmholtz free energy
is obtained as a function of the order parameter and total pair momentum. We
also analyze the extent of the phase separation region determined by the
minimal free energy of a normal and a BCS mixture. The polarization \textit{%
vs} temperature phase diagram is calculated and compared with previous
results for a 2D system \cite{KMF}. In Section IV we derive a Bethe-Salpeter
equation for two-body amplitudes assuming a generalized random-phase
approximation. The collective excitations are obtained via the vanishing of
a secular $4\times 4$ determinant. Finally, our conclusions are presented in
Section V.

\section{Hubbard model in a cubic optical lattice}

The Hamiltonian of a two-component Fermi gas under an attractive contact
interaction $v(\boldsymbol{r}-\boldsymbol{r}^{\prime })=v_{0}\delta (%
\boldsymbol{r}-\boldsymbol{r}^{\prime })$ is given by \hspace{-1.3cm}%
\begin{eqnarray}
H =\sum_{\sigma }\int d\boldsymbol{r}\hat{\Psi}_{\sigma }^{\dag }(%
\boldsymbol{r})\left[ -\frac{\hbar ^{2}\nabla ^{2}}{2m}+V_{\sigma }(%
\boldsymbol{r})-\mu _{\sigma }\right] \hat{\Psi}_{\sigma }(\boldsymbol{r}) &&
\notag \\
+v_{0}\int \int d\boldsymbol{r}d\boldsymbol{r}^{\prime }\hat{\Psi}_{\sigma
_{1}}^{\dag }(\boldsymbol{r})\hat{\Psi}_{\sigma _{2}}^{\dag }(\boldsymbol{r}%
^{\prime })\delta (\boldsymbol{r}-\boldsymbol{r}^{\prime })\hat{\Psi}%
_{\sigma _{2}}(\boldsymbol{r}^{\prime })\hat{\Psi}_{\sigma _{1}}(\boldsymbol{%
r}),&&  \label{hamil}
\end{eqnarray}%
where $\hat{\Psi}_{\sigma }^{\dag }(\boldsymbol{r})$ and $\hat{\Psi}_{\sigma
}(\boldsymbol{r})$ are fermionic creation and annihilation field operators
of component $\sigma $, respectively; $\mu _{\sigma }$ is the chemical
potential for each component $|\!\!\uparrow >$ or $|\!\!\downarrow >$, and
the lattice periodic potential is%
\begin{equation}
V_{\sigma }(\boldsymbol{r})=\gamma _{\sigma ,x}\sin ^{2}kx+\gamma _{\sigma
,y}\sin ^{2}ky+\gamma _{\sigma ,z}\sin ^{2}kz,  \label{perpot}
\end{equation}%
where $k=\pi /a$ with $a=\lambda /2$, the lattice constant and $\lambda $ is
the laser wavelength.

We will assume that the optical-lattice potential strengths $\gamma _{\sigma
,\nu }$ ($\nu =x,y\,\ $or$\ \,z$) are sufficiently deep to consider that
lattice-site tunneling occurs only between nearest neighbors. Then, the
field operators can be expanded as%
\begin{equation*}
\hat{\Psi}_{\sigma }(\boldsymbol{r})=\sum_{i}\psi _{i,\sigma }(\boldsymbol{r}%
)\hat{c}_{i,\sigma },
\end{equation*}%
where $\psi _{i,\sigma }(\boldsymbol{r})$ are one-particle wave functions
localized at site $i$, and the Fermi operator $\hat{c}_{i,\sigma }^{\dag }$ (%
$\hat{c}_{i,\sigma }$) creates (destroys) an atom in site $i.$ Under these
assumptions, the Hamiltonian in (\ref{hamil}) reduces to the single-band
attractive Hubbard model, 
\begin{eqnarray}
H &=&-J_{x}\sum_{\left\langle i,j\right\rangle _{x},\sigma }\hat{c}%
_{i,\sigma }^{\dag }\hat{c}_{j,\sigma }-J_{y}\sum_{\left\langle
i,j\right\rangle _{y},\sigma }\hat{c}_{i,\sigma }^{\dag }\hat{c}_{j,\sigma }
\notag \\
&& - J_{z}\sum_{\left\langle i,j\right\rangle _{z},\sigma }\hat{c}_{i,\sigma
}^{\dag }\hat{c}_{j,\sigma } -\sum\limits_{i}\left( \mu _{\uparrow }^{\dag }%
\hat{c}_{i,\uparrow }^{\dag }\hat{c}_{i,\uparrow }+\mu _{\downarrow }\hat{c}%
_{i,\downarrow }^{\dag }\hat{c}_{i,\downarrow }\right)  \notag \\
&& +U\sum\limits_{i}\hat{c}_{i,\uparrow }^{\dag }\hat{c}_{i,\downarrow
}^{\dag }\hat{c}_{i,\downarrow }\hat{c}_{i,\uparrow },  \label{Hub}
\end{eqnarray}%
where $J_{\nu }$ is the tunneling strength of the atoms between
nearest-neighbor sites in the $\nu $-direction and $U$ is the on-site
attractive interaction strength. On the BCS side, the Hubbard parameter $U$
is negative, but in what follows $U$ denotes its absolute value and is given
by%
\begin{equation}
U=v_{0}\int d\boldsymbol{r}\left\vert \psi _{i,\uparrow }(\boldsymbol{r}%
)\right\vert ^{2}\left\vert \psi _{i,\downarrow }(\boldsymbol{r})\right\vert
^{2}.  \label{U}
\end{equation}

We assume a system with a total number of atoms $M=M_{\uparrow
}+M_{\downarrow }$ distributed along $N$ sites of the optical-lattice
potential (\ref{perpot}). In the mean-field approximation, the pair
interaction term in Eq. (\ref{Hub}) is replaced by%
\begin{eqnarray}
&& U\sum\limits_{i}\hat{c}_{i,\uparrow }^{\dag }\hat{c}_{i,\downarrow
}^{\dag } \hat{c}_{i,\downarrow }\hat{c}_{i,\uparrow } \simeq
U\sum\limits_{i} \left( \left\langle \hat{c}_{i,\uparrow }^{\dag }\hat{c}%
_{i,\downarrow }^{\dag }\right\rangle \hat{c}_{i,\downarrow }\hat{c}%
_{i,\uparrow } \right.  \notag \\
&& \left. +\hat{c} _{i,\uparrow }^{\dag }\hat{c}_{i,\downarrow }^{\dag
}\left\langle \hat{c}_{i,\downarrow }\hat{c}_{i,\uparrow }\right\rangle
-\left\langle \hat{c}_{i,\uparrow }^{\dag }\hat{c}_{i,\downarrow }^{\dag
}\right\rangle \left\langle \hat{c}_{i,\downarrow }\hat{c}_{i,\uparrow
}\right\rangle \right).  \label{mf}
\end{eqnarray}

The order parameter $\Delta _{i}=U\left\langle \hat{c}_{i,\downarrow }\hat{c}%
_{i,\uparrow }\right\rangle $ of the FFLO states is assumed to vary as a
single plane wave, $\Delta _{i}=\Delta \exp \left( 2\imath \boldsymbol{q}%
\cdot \boldsymbol{r}_{i}\right) $, where $\boldsymbol{q}$ is the pair
center-of-mass momentum and $\boldsymbol{r}_{i}$ the coordinate of site $i.$
These states are expected to occur on the BCS side of a Feshbach resonance
where the effective attractive interaction between fermion atoms leads to
BCS-type pairing. The tight-binding lattice dispersion energy is $\xi
_{\uparrow ,\downarrow }(\boldsymbol{k})=2J\left( 1-\sum_{\nu }\cos k_{\nu
}a\right) -\mu _{\uparrow ,\downarrow }.$ In our calculations we use $%
\lambda =1030$ nm and equal tunneling strengths $J_{\nu }=J$ to obtain the
following mean-field Hamiltonian, 
\begin{eqnarray}
&& \hspace{-0.80cm} H=\frac{1}{N}\sum\limits_{\boldsymbol{k}}\left[ \xi
_{\uparrow }(\boldsymbol{k})\hat{c}_{\boldsymbol{k},\uparrow }^{\dag }\hat{c}%
_{\boldsymbol{k},\uparrow }+\xi _{\downarrow }(\boldsymbol{k})\hat{c}_{%
\boldsymbol{k},\downarrow }^{\dag }\hat{c}_{\boldsymbol{k},\downarrow }
\right.  \notag \\
&& \hspace{-0.5cm} + \! \! \left. \Delta \hat{c}_{\boldsymbol{k+q},\uparrow
}^{\dag }\hat{c}_{\boldsymbol{-k+q},\downarrow }^{\dag } + \Delta ^{\ast }%
\hat{c}_{\boldsymbol{-k+q},\downarrow }\hat{c}_{\boldsymbol{k+q},\uparrow }+%
\frac{\left\vert \Delta \right\vert ^{2}}{U} \right],  \label{mfmom}
\end{eqnarray}
which can be diagonalized using a Bogoliubov transformation \cite{Kop1},%
\begin{equation}
\left( 
\begin{array}{c}
\hat{c}_{\boldsymbol{k+q},\uparrow } \\ 
\hat{c}_{-\boldsymbol{k+q},\downarrow }^{\dag }%
\end{array}%
\right) =\left( 
\begin{array}{cc}
u_{\boldsymbol{q}}(\boldsymbol{k}) & v_{\boldsymbol{q}}(\boldsymbol{k}) \\ 
-v_{\boldsymbol{q}}(\boldsymbol{k}) & u_{\boldsymbol{q}}(\boldsymbol{k})%
\end{array}%
\right) \left( 
\begin{array}{c}
\hat{d}_{\boldsymbol{k,q},\uparrow } \\ 
\hat{d}_{-\boldsymbol{k,q},\downarrow }^{\dag }%
\end{array}%
\right) .  \label{Bogtr}
\end{equation}%
The coefficients $u_{\boldsymbol{q}},v_{\boldsymbol{q}}$ are given by%
\begin{equation}
u_{\boldsymbol{q}}(\boldsymbol{k})=\sqrt{\frac{1}{2}\left[ 1+\frac{\chi _{%
\boldsymbol{q}}(\boldsymbol{k})}{E_{\boldsymbol{q}}(\boldsymbol{k})}\right] }%
,\text{ \ }v_{\boldsymbol{q}}(\boldsymbol{k})=\sqrt{\frac{1}{2}\left[ 1-%
\frac{\chi _{\boldsymbol{q}}(\boldsymbol{k})}{E_{\boldsymbol{q}}(\boldsymbol{%
k})}\right] },  \label{uv}
\end{equation}%
where%
\begin{eqnarray}
\chi _{\boldsymbol{q}}(\boldsymbol{k}) &=&\frac{1}{2}\left[ \xi _{\uparrow }(%
\boldsymbol{k+q})+\xi _{\downarrow }(\boldsymbol{q-k})\right] ,  \notag \\
E_{\boldsymbol{q}}(\boldsymbol{k}) &=&\sqrt{\chi _{\boldsymbol{q}}^{2}(%
\boldsymbol{k})+\Delta ^{2}}  \label{etacsi}
\end{eqnarray}

\section{Thermodynamic potential and phase diagrams}

In the mean-field approximation, the momentum-space, single-particle Green
function is a $2\times 2$ matrix given by%
\begin{equation*}
\widehat{G}=\left( 
\begin{array}{cc}
G_{\boldsymbol{q}}^{\uparrow \uparrow } & G_{\boldsymbol{q}}^{\uparrow
\downarrow } \\ 
G_{\boldsymbol{q}}^{\downarrow \uparrow } & G_{\boldsymbol{q}}^{\downarrow
\downarrow }%
\end{array}%
\right)
\end{equation*}%
where%
\begin{align}
G_{\boldsymbol{q}}^{\uparrow \uparrow }(\boldsymbol{k},\imath \omega _{m})& =%
\frac{u_{\boldsymbol{q}}(\boldsymbol{k})^{2}}{\imath \hbar \omega
_{m}-\omega _{+}(\boldsymbol{k},\boldsymbol{q})}+\frac{v_{\boldsymbol{q}}(%
\boldsymbol{k})^{2}}{\imath \hbar \omega _{m}+\omega _{-}(\boldsymbol{k},%
\boldsymbol{q})},  \notag \\
G_{\boldsymbol{q}}^{\downarrow \downarrow }(\boldsymbol{k},\imath \omega
_{m})& =\frac{v_{\boldsymbol{q}}(\boldsymbol{k})^{2}}{\imath \hbar \omega
_{m}-\omega _{+}(\boldsymbol{k},\boldsymbol{q})}+\frac{u_{\boldsymbol{q}}(%
\boldsymbol{k})^{2}}{\imath \hbar \omega _{m}+\omega _{-}(\boldsymbol{k},%
\boldsymbol{q})},  \notag \\
G_{\boldsymbol{q}}^{\uparrow \downarrow }(\boldsymbol{k},\imath \omega
_{m})& =G_{\boldsymbol{q}}^{\downarrow \uparrow }(\boldsymbol{k},\imath
\omega _{m})=u_{\boldsymbol{q}}(\boldsymbol{k)}v_{\boldsymbol{q}}(%
\boldsymbol{k)} \times  \notag \\
& \hspace{-0.50cm} \left[ \frac{1}{\imath \hbar \omega _{m}-\omega _{+}(%
\boldsymbol{k},\boldsymbol{q})}-\frac{1}{\imath \hbar \omega _{m}+\omega
_{-}(\boldsymbol{k},\boldsymbol{q})}\right] .  \label{gf}
\end{align}

The Matsubara frequencies are $\omega _{m}=\pi k_{B}T(2m+1)/\hbar $ with $%
m=0,$ $\pm 1,$ $\pm$ 2,...; $T$ is the temperature, and $k_{B}$ the Boltzmann
constant. The one-particle excitations in a mean-field approximation $\omega
_{\pm }$ are coherent combinations of electronlike $\omega _{+}(\boldsymbol{k%
},\boldsymbol{q})=E_{\boldsymbol{q}}(\boldsymbol{k})+\eta _{\boldsymbol{q}}(%
\boldsymbol{k})$ and holelike $\omega _{-}(\boldsymbol{k},\boldsymbol{q})=E_{%
\boldsymbol{q}}(\boldsymbol{k})-\eta _{\boldsymbol{q}}(\boldsymbol{k})$
excitations, where $\eta _{\boldsymbol{q}}(\boldsymbol{k}) =\frac{1}{2}\left[
\xi _{\uparrow }(\boldsymbol{k+q})-\xi _{\downarrow }(\boldsymbol{q-k})%
\right]$. The thermodynamic potential at temperature $T$ in a mean field
approximation can be evaluated from the grand canonical partition function $%
Z $ of an ensemble of quasiparticles with energy $\omega _{\pm }(\boldsymbol{%
k},\boldsymbol{q})$ given by \cite{Kop1} 
\begin{eqnarray}
Z=\prod\limits_{\boldsymbol{k}}\left( 1+e^{-\beta \omega _{+}(\boldsymbol{k},%
\boldsymbol{q})/N}\right) && \hspace{-0.40cm} \left( 1+e^{\beta \omega _{-}(%
\boldsymbol{k},\boldsymbol{q})/N}\right) \times  \notag \\
&& \hspace{-0.3cm} e^{-\frac{\beta }{N}\left( \chi _{\boldsymbol{q}}(%
\boldsymbol{k})+\frac{\left\vert \Delta \right\vert ^{2}}{U}\right) },
\label{part}
\end{eqnarray}%
where $\beta =1/k_{B}T.$ The thermodynamic potential $\Omega =-\frac{1}{%
\beta }\ln Z$ is therefore, 
\begin{eqnarray}
\Omega &=&\frac{1}{N}\sum_{\boldsymbol{k}}\left[ \chi _{\boldsymbol{q}}(%
\boldsymbol{k})+\omega _{-}(\boldsymbol{k},\boldsymbol{q})+\frac{\Delta ^{2}%
}{U}\right]  \notag \\
&& \hspace{-1.5cm} -\frac{1}{\beta }\sum_{\boldsymbol{k}}\left[ \ln \left(
1+e^{-\beta \omega _{+}(\boldsymbol{k},\boldsymbol{q})})+\ln (1+e^{\beta
\omega _{-}(\boldsymbol{k},\boldsymbol{q})}\right) \right] .
\label{thermpot}
\end{eqnarray}

From (\ref{uv}), the parameter $\Delta =\frac{U}{N}\sum_{\boldsymbol{k}%
}\left\langle \hat{c}_{\boldsymbol{-k+q},\downarrow }\hat{c}_{\boldsymbol{k+q%
},\uparrow }\right\rangle $ satisfies the gap equation at zero temperature%
\begin{equation}
1=\frac{U}{N}\sum_{\boldsymbol{k}}\frac{1}{2E_{\boldsymbol{q}}(\boldsymbol{k}%
)}.  \label{gapeq}
\end{equation}

If we consider an imbalanced system with fixed chemical potentials, $\mu
_{\uparrow ,\downarrow }$, the minima of $\Omega (\Delta ,\boldsymbol{q},\mu
_{\uparrow },\mu _{\downarrow },T)$ with respect to $\Delta ,\boldsymbol{q}%
,\mu _{\uparrow },\mu _{\downarrow }$ define the possible stable phases of
this system as a function of temperature. However, recent experiments \cite%
{Ketterle2, Ketterle} deal with the more realistic situation in which the number of
particles of each kind is fixed. In the latter case, the relevant
thermodynamic potential is the Helmholtz free energy $F(\Delta ,\boldsymbol{q%
},f_{\uparrow },f_{\downarrow },T)=\Omega +\mu _{\uparrow }f_{\uparrow }+\mu
_{\downarrow }f_{\downarrow }$. Without loss of generality we set $%
\boldsymbol{q}=(q_{x},0,0),$ i.e., in the $x$-direction, and minimize the
Helmholtz free energy $F(\Delta ,q_{x},f_{\uparrow },f_{\downarrow },T)$
with respect to $\mu _{\uparrow }$, $\mu _{\downarrow }$, $\Delta $ and $%
q_{x}$, where $f_{\uparrow ,\downarrow }\equiv M_{\uparrow ,\downarrow }/N$.
As a result, we obtain a set of four equations, namely the number and gap
equations, as well as the equation for $q_{x}$: 
\begin{align}
& f_{\uparrow }=\frac{1}{N}\sum_{\boldsymbol{k}}\left[ u_{\boldsymbol{q}%
}^{2}(\boldsymbol{k})f(\omega _{+}(\boldsymbol{k},\boldsymbol{q}))+v_{%
\boldsymbol{q}}^{2}(\boldsymbol{k})f(-\omega _{-}(\boldsymbol{k},\boldsymbol{%
q}))\right] ,  \notag \\
& f_{\downarrow }=\frac{1}{N}\sum_{\boldsymbol{k}}\left[ u_{\boldsymbol{q}%
}^{2}(\boldsymbol{k})f(\omega _{-}(\boldsymbol{k},\boldsymbol{q}))+v_{%
\boldsymbol{q}}^{2}(\boldsymbol{k})f(-\omega _{+}(\boldsymbol{k},\boldsymbol{%
q}))\right] ,  \notag \\
&1 =\frac{U}{N}\sum_{\boldsymbol{k}}\frac{1-f(\omega _{-}(\boldsymbol{k},%
\boldsymbol{q}))-f(\omega _{+}(\boldsymbol{k},\boldsymbol{q}))}{2E_{%
\boldsymbol{q}}(\boldsymbol{k})}  \notag \\
&0 =\frac{1}{N}\sum_{\boldsymbol{k}} \left\{ \frac{\partial \eta _{%
\boldsymbol{q}}(\boldsymbol{k})}{\partial q_{x}}\left[ f(\omega _{+}(%
\boldsymbol{k},\boldsymbol{q}))-f(\omega _{-}(\boldsymbol{k},\boldsymbol{q}))%
\right] +\frac{\partial \chi _{\boldsymbol{q}}(\boldsymbol{k})}{\partial
q_{x}} \right.  \notag \\
&\! \times \! \left. \left[ 1-\frac{\chi _{\boldsymbol{q}}(\boldsymbol{k})}{%
E_{\boldsymbol{q}}(\boldsymbol{k})}\left[ 1-f(\omega _{+}(\boldsymbol{k},%
\boldsymbol{q}))-f(\omega _{-}(\boldsymbol{k},\boldsymbol{q}))\right] \right]%
\right\} ,  \label{foureqs}
\end{align}%
where $f(\omega _{\pm }(\boldsymbol{k},\boldsymbol{q}))=\left\langle \hat{d}%
_{-\boldsymbol{k,q},\uparrow }^{\dag }\hat{d}_{\boldsymbol{k,q},\uparrow
}\right\rangle$ is the Fermi distribution $\left[ \exp \left( \beta \omega
_{\pm }(\boldsymbol{k}, \boldsymbol{q})\right) +1\right] ^{-1}$.

The existence of mixed phases of normal state and superfluid has been
reported in several analysis \cite{Kop}$^{,}$ \cite{Cald}. It arises when a fraction 
of the fermions are forming Cooper pairs in a BCS, Sarma or FFLO phase while the remaining
(imbalanced) atoms are in the normal phase. Here, we only consider a configuration in which a fraction $(1-x)$ are in the BCS phase which requires equal numbers
of $|\!\!\uparrow >$ and $|\!\!\downarrow >$ states with opposite momenta.
The free energy in this mixed or phase separation (PS) state is%
\begin{equation}
F_{PS}=xF_{N}+(1-x)F_{BCS},  \label{psfree}
\end{equation}%
where%
\begin{equation}
F_{BCS}=\Omega _{BCS}+\mu (1-x)\tilde{f},  \label{BCSfree}
\end{equation}%
and $(1-x)\tilde{f}$ is the filling-factor fraction of fermions in the BCS
state, and $\mu $ is the chemical potential with similar expressions for the
filling-factor fraction in the normal phase given by%
\begin{equation}
F_{N}=\Omega _{N}+\mu _{\uparrow }\left( f_{\uparrow }-(1-x)\tilde{f}\right)
+\mu _{\downarrow }\left( f_{\downarrow }-(1-x)\tilde{f}\right) ,
\label{Fnormal}
\end{equation}%
where the thermodynamic potential in the normal phase is%
\begin{equation}
\Omega _{N}=-\frac{1}{\beta }\sum_{\mathbf{k}}\left\{ \ln \left[
(1+e^{-\beta \Omega _{\uparrow }})(1+e^{-\beta \Omega _{\downarrow }})\right]
\right\} .  \label{omegaN}
\end{equation}

The free energy is now also a function of $x,$ $\tilde{f}$ and $\mu $, $i.e.$, 
 $F_{PS}=F_{PS}(x,$ $\tilde{f}$ $,\mu ,\mu _{\uparrow },\mu _{\downarrow
},\Delta ).$ The minimum of $F_{PS}$ with respect to variations in the
normal fraction $x$ or $\tilde{f}$ results in the following two additional
relations%
\begin{eqnarray}
\tilde{f}(\mu _{\uparrow }+\mu _{\downarrow }) &=&\Omega _{N}-\Omega
_{BCS}+\mu _{\uparrow }f_{\uparrow }+\mu _{\downarrow }f_{\downarrow }, 
\notag \\
\mu (1-x) &=&x(\mu _{\uparrow }+\mu _{\downarrow }),  \label{psrelations}
\end{eqnarray}%
which together with Eqs. (\ref{foureqs}) provide a system of six equations
that define the equilibrium values of the thermodynamic variables. 
\begin{figure}[tbh]
\centerline{\epsfig{file=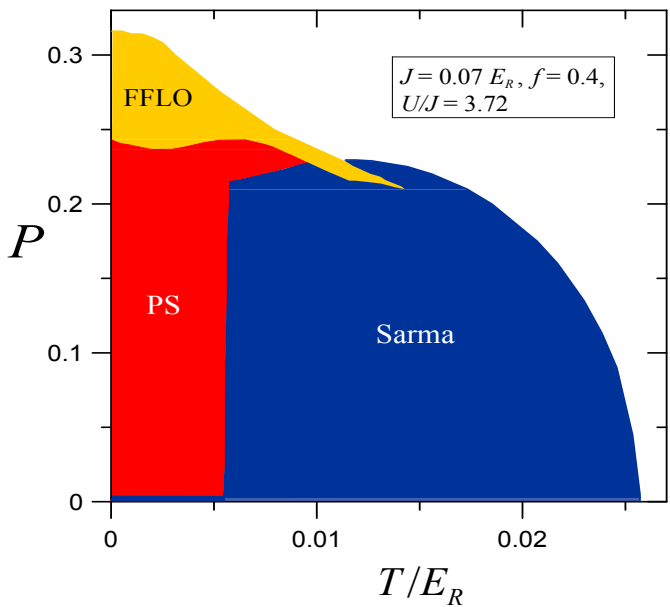,height=3.2in,width=3.50in}}
\caption{(Color online). Phase diagram of a polarized $^{6}$Li gas in a 3D
optical lattice with $\protect\lambda =1010$ nm and filling factor $f=0.4.$
FFLO (yellow), phase separation (red), Sarma (blue). The Hubbard parameters
are $J=0.07\ E_{R}$ and the attractive on-site attractive interaction is $%
U/J=3.72$.}
\label{fig:PD3D1}
\end{figure}
Since at a finite temperature the FFLO, Sarma, PS and normal states compete
with each other, we have calculated the regions in the $P$ $vs$ $T$ plane
that minimize the free energy. Here, $P$ is the polarization defined by%
\begin{equation}
P=\frac{f_{\uparrow }-f_{\downarrow }}{f_{\uparrow }+f_{\downarrow }}.
\label{pol}
\end{equation}

In Fig. \ref{fig:PD3D1} we exhibit the phase diagram of a three-dimensional
imbalanced system for a total filling factor $f=f_{\uparrow }+f_{\downarrow
}=0.4$, $J=0.07\ E_{R},$ and $U/J=3.72$, where $E_{R}=\hbar ^{2}(2\pi
/\lambda )^{2}/2m$ is the recoil energy. We first choose these values for
the parameters in order to compare with the results of reference \cite{Kop}.
At low temperatures and polarization $P$ $\geq $ $0.25$ the FFLO states are
shown to be more stable than the Sarma phase, where the latter is
characterized by $\boldsymbol{q}=0,$ $\Delta \neq 0$ and $P\neq 0.$ The
stability phase region of the FFLO states extends to temperatures up to $%
k_{B}T/E_{R}\approx 0.015$ albeit over a narrower polarization interval
compared to that obtained in \cite{Kop}. This difference may be due to the
definition of the phase separation phase given in Eq. (\ref{psfree}). In the
BCS phase, $\boldsymbol{q}=0$ and the number of particles of each species is
the same, i.e., $P=0.$ There is also the mixed phase region composed of
normal and superfluid states where a fraction of the fluid is in the normal
phase while the remaining fraction is in the BCS phase \cite{Cald}.

\begin{figure}[tbh]
\centerline{\epsfig{file=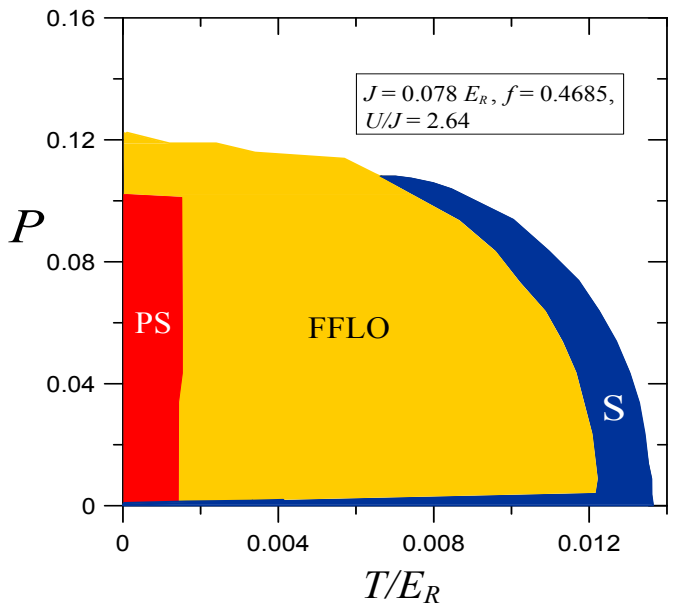,height=3.2in,width=3.50in}}
\caption{(Color online). FFLO (yellow), phase separation (red) and Sarma
(blue) phases of a polarized $^{6}$Li gas in a 3D optical lattice with $%
\protect\lambda =1030$ nm and filling factor $f=0.4685.$ The Hubbard
parameters are $J=0.078\ E_{R}$ and the attractive on-site attractive
interaction is $U/J=2.64.$}
\label{fig:PhD3D2}
\end{figure}

In Fig. \ref{fig:PhD3D2} we show the phase diagram for the same system but
with a weaker attraction term $U/J=2.64$ and $f=0.4685$. This value for the
on-site attraction coincides with our previous results \cite{KMF} in 2D as
we are interested in analyzing the effects of dimensionality on these
systems. A decrease in $U$ and a slight increase in $J$ enhances the hopping
between nearest neighboring sites. The overall effect is to expand the
stability region of the FFLO phase in relation to the Sarma states compared
to the phase diagram of Fig. \ref{fig:PD3D1}. The largest polarization that
the system can support before it becomes a normal fluid is $P=0.124.$ In
this case, the FFLO states lower the system free energy over quite a large
phase region compared to the corresponding Sarma states at low temperatures.
As the temperature increases, a sliver of Sarma states provides the minimum
of the free energy. If the temperature is increased even further, the normal
polarized Fermi gas becomes the energetically favored phase. Here, the
phase-separation region and the FFLO phase dominate over the Sarma states.

\begin{figure}[tbh]
\centerline{\epsfig{file=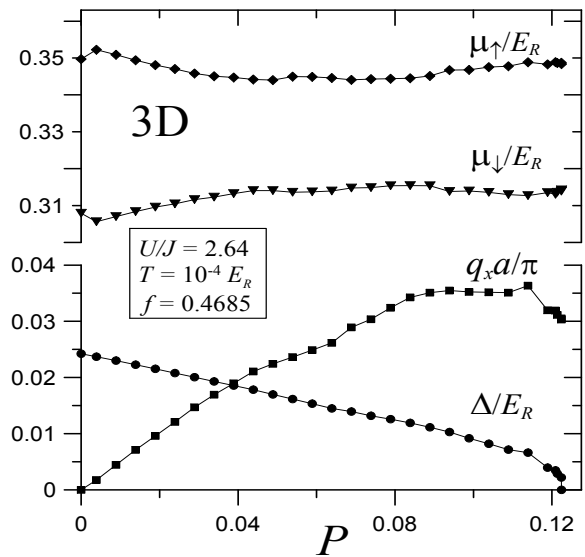,height=3.0in,width=3.20in}}
\caption{Chemical potentials, pair momentum and gap for an imbalanced
fermion gas loaded in a 3D optical lattice at $k_{B}T=10^{-4} \ E_{R}$}
\label{fig:mu-gap3D}
\end{figure}

\begin{figure}[tbh]
\centerline{\epsfig{file=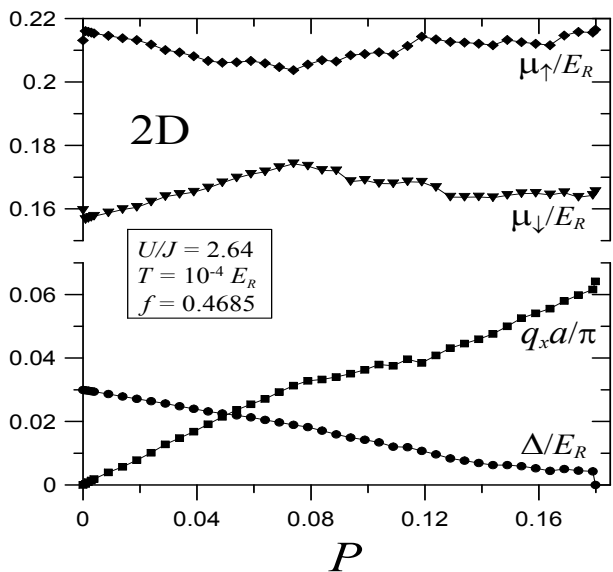,height=3.0in,width=3.20in}}
\caption{Chemical potentials, pair momentum and gap for an imbalanced
fermion gas loaded in a 2D optical lattice with $k_{B}T=10^{-4} \ E_{R}.$}
\label{fig:mu-gap2D}
\end{figure}

In Figs. \ref{fig:mu-gap3D} and \ref{fig:mu-gap2D} we exhibit the variation
with the polarization of the chemical potential of each species, the
pair-momentum and the gap at a fixed temperature $k_{B}T=10^{-4}\ E_{R}.$
Figure \ref{fig:mu-gap3D} shows the results for a 3D system where it remains
as a FFLO superfluid up to $P\simeq 0.124.$ At this value of the
polarization the gap vanishes and therefore it enters a normal phase. In
contrast, Fig. \ref{fig:mu-gap2D} shows the behavior of these quantities in
a 2D system with the same parameters $U,$ $J$ and $f$. It is interesting to
note that even though the variation of $\mu _{\uparrow }$, $\mu _{\downarrow
}$, $\Delta $ and $q_{x}$ follows the same trend as in the 3D case, the
system remains a FFLO superfluid up to a somewhat higher value of the
polarization, $P\simeq 0.18$ in the 2D regime.

\section{Collective states}

Unlike the population-balanced systems, for which the spectrum of the
collective excitations has been obtained by linearizing the
Anderson-Rickayzen equations \cite{BR}, by the Kadanoff and Baym approach 
\cite{CG} and by the Bethe-Salpeter (BS) formalism \cite{ZK}, the FFLO
collective modes have been studied in: (i) a 1D population-unbalanced
trapped system \cite{HT} by using the linear response of the equilibrium
system by supplementing the Bogoliubov--de Gennes (BdG) equations with a
self-consistent random phase approximation; (ii) a 1D superconductor \cite%
{KS} by transforming slow deformations of the order parameter into small
corrections to the BdG Hamiltonian; and (iii) a cold-atom rotated system 
\cite{AMD} by locating the poles of the many-body scattering function. Here,
we present a theory that goes beyond the mean-field approaches to find the
spectrum of the collective excitations in the presence of FFLO phase by
solving the BS equations for this spectrum in the general random phase
approximation (GRPA) in a 3D optical lattice \cite{KMF}.

The spectrum of the collective modes can be obtained from the poles of the
two-particle Green's function $K(1,2;3,4),$ where we use the compact
notation $1=\{\sigma _{1},\boldsymbol{r}_{1},t_{1}\},$ $2=\{\sigma _{2},%
\boldsymbol{r}_{2},t_{2}\},...$ with $\sigma _{i}$ denoting the spin
variables, $\boldsymbol{r}_{i}$ the vector for lattice site $i$, and $t_{i}$%
, the time variable. $K$ satisfies the following Dyson equation:%
\begin{equation}
K=K_{0}+K_{0}IK,  \label{BSdyson}
\end{equation}%
where $K_{0}(1,2;3,4)$ is the two-particle free propagator which is defined
by a pair of fully dressed single-particle Green%
\'{}%
s function,%
\begin{equation*}
K_{0}(1,2;3,4)=G(1;3)G(4;2).
\end{equation*}%
The interaction kernel $I$ is given by the functional derivatives of the
mass operator $\Sigma (1;2)=\Sigma _{D}(1;2)+\Sigma _{E}(1;2)$ obtained from
the direct (or Fock) and exchange (or Hartree) parts, $I=\frac{\delta \Sigma 
}{\delta G}=\frac{\delta \Sigma _{D}}{\delta G}+\frac{\delta \Sigma _{E}}{%
\delta G}$. The Dyson equation for $G$ is%
\begin{equation}
\widehat{G}=G_{0}+G_{0}\Sigma \widehat{G}  \label{Gdyson}
\end{equation}%
and therefore, the equation for the two-particle Green's function (\ref%
{BSdyson}) must be solved self-consistently with (\ref{Gdyson}). Since we
are interested in the collective energy $\omega (\boldsymbol{Q)}$ and momentum $\boldsymbol{Q}$ excitations which are
given by the poles of the two-particle fully dressed Green's function, we
write the latter using the spectral representation%
\begin{eqnarray}
&&K(1,2;3,4)=\sum_{\omega _{p}}e^{-i\omega _{p}(u_{1}-u_{3})}   \notag \\
&&\hspace{-1cm}\times \frac{\Phi _{\boldsymbol{Q;}\sigma _{1},\sigma _{2}}(%
\boldsymbol{r}_{i_{1}},\boldsymbol{r}_{i_{2}};u_{1}-u_{2})\Phi _{\boldsymbol{%
Q}}^{\ast }(\boldsymbol{r}_{i_{3}},\boldsymbol{r}_{i_{4}};u_{3}-u_{4})}{%
i\omega _{p}-\omega (\boldsymbol{Q)}},  \label{spectr}
\end{eqnarray}%
where $\Phi _{\boldsymbol{Q;}\sigma _{1},\sigma _{2}}(\boldsymbol{r}_{i_{1}},%
\boldsymbol{r}_{i_{2}};u_{2}-u_{1})$ are the BS amplitudes%
\begin{eqnarray*}
\Phi _{\boldsymbol{Q;}\sigma _{1},\sigma _{2}}(\boldsymbol{r}_{i_{1}},%
\boldsymbol{r}_{i_{2}};u_{2}-u_{1}) &&\hspace{-0.3cm}=e^{i\boldsymbol{Q}%
\cdot (\boldsymbol{r}_{i_{1}}+\boldsymbol{r}_{i_{2}})/2}\times \\
&&\phi _{\boldsymbol{Q;}\sigma _{1},\sigma _{2}}(\boldsymbol{r}_{i_{1}}-%
\boldsymbol{r}_{i_{2}};u_{1}-u_{2}).
\end{eqnarray*}%
In the momentum-space representation and with equal time components, $%
u_{1}=u_{2}$ we have%
\begin{equation}
\phi _{\boldsymbol{Q;}\sigma _{1},\sigma _{2}}(\boldsymbol{r}_{i_{1}}-%
\boldsymbol{r}_{i_{2}};0)=\frac{1}{N}\sum\limits_{\boldsymbol{k}}e^{ik\cdot (%
\boldsymbol{r}_{i_{1}}-\boldsymbol{r}_{i_{2}})}\phi _{\sigma _{1},\sigma
_{2}}(\boldsymbol{k},\boldsymbol{Q}).  \label{phiexp}
\end{equation}

It is widely accepted that the generalized random phase is a good
approximation for the collective excitations in a weak-coupling regime, and
therefore, it can be used to separate the solutions of the Dyson and the
Bethe-Salpeter equations. In this approximation, the single-particle
excitations are replaced with those obtained by diagonalizing the
Hartree-Fock (HF) Hamiltonian; while the collective modes are obtained by
solving the BS equation in which the single-particle Green's functions are
calculated in HF approximation, and the BS kernel is obtained by summing
ladder and bubble diagrams.

Inserting expansion (\ref{phiexp}) in Eq. (\ref{BSdyson}) using (\ref{spectr}%
) 
\begin{widetext}
\begin{equation*}
\phi _{\boldsymbol{q,}\sigma _{1},\sigma _{2}}(\boldsymbol{k},\boldsymbol{Q}%
)=\sum\limits_{\sigma _{3},\sigma _{4},\sigma _{1}^{\prime },\sigma
_{2}^{\prime }}\sum\limits_{\imath \omega _{m}}G_{\boldsymbol{q}}^{\sigma
_{1}\sigma _{3}}(\boldsymbol{k}+\boldsymbol{Q},\imath \omega _{m}+\omega (%
\boldsymbol{Q}))G_{\boldsymbol{q}}^{\sigma _{4}\sigma _{2}}(\boldsymbol{k}%
,\imath \omega _{m})I_{\sigma _{3},\sigma _{4},\sigma _{1}^{\prime },\sigma
_{2}^{\prime }}\sum\limits_{\boldsymbol{p}}\phi _{\boldsymbol{q,}\sigma
_{1}^{\prime },\sigma _{2}^{\prime }}(\boldsymbol{p},\boldsymbol{Q}),
\end{equation*}%
\end{widetext}where the kernel represents the direct and exchange
interactions: 
\begin{eqnarray}
I_{\sigma _{1},\sigma _{2},\sigma _{3},\sigma _{4}} &=&I_{\sigma _{1},\sigma
_{2},\sigma _{3},\sigma _{4}}^{d}+I_{\sigma _{1},\sigma _{2},\sigma
_{3},\sigma _{4}}^{exch},  \notag \\
I_{\sigma _{1},\sigma _{2},\sigma _{3},\sigma _{4}}^{d} &=&-U\delta _{\sigma
_{1},\sigma _{3}}\delta _{\sigma _{2},\sigma _{4}},  \notag \\
I_{\sigma _{1},\sigma _{2},\sigma _{3},\sigma _{4}}^{exch} &=&U\delta
_{\sigma _{1},\sigma _{2}}\delta _{\sigma _{3},\sigma _{4}}  \label{Iker}
\end{eqnarray}

We now introduce the compact notation $\hat{\phi}_{\boldsymbol{q}}(%
\boldsymbol{k},\boldsymbol{Q})=\left[ \phi _{\boldsymbol{q,}\downarrow
,\uparrow }(\boldsymbol{k},\boldsymbol{Q}),\phi _{\boldsymbol{q,}\uparrow
,\downarrow }(\boldsymbol{k},\boldsymbol{Q}),\phi _{\boldsymbol{q,}\uparrow
,\uparrow }(\boldsymbol{k},\boldsymbol{Q}),\phi _{\boldsymbol{q,}\downarrow
,\downarrow }(\boldsymbol{k},\boldsymbol{Q})\right] ^{T}$ (where $T$ means
the transpose vector). Then, the equation for the BS amplitudes becomes%
\begin{equation}
\hat{\phi}_{\boldsymbol{q}}(\boldsymbol{k},\boldsymbol{Q})=-U\hat{D}\sum_{%
\mathbf{p}}\hat{\phi}_{\boldsymbol{q}}(\mathbf{p},\boldsymbol{Q})+U\hat{M}%
\sum_{\mathbf{p}}\hat{\phi}_{\boldsymbol{q}}(\mathbf{p},\boldsymbol{Q}).
\label{BSEampl}
\end{equation}%
Here, $U\widehat{D}$ and $U\widehat{M}$ represent the direct and exchange
interactions, respectively:%
\begin{widetext}
\begin{equation}
\hspace{-1.50cm} \widehat{D}=\left(
\begin{array}{cccc}
K_\textbf{q}^{\left(\downarrow,\downarrow,\uparrow,\uparrow\right)}(\textbf{k},\textbf{Q},\imath\omega_p),&
K_\textbf{q}^{\left(\downarrow,\uparrow,\downarrow,\uparrow\right)}(\textbf{k},\textbf{Q},\imath\omega_p)&0&0\\
K_\textbf{q}^{\left(\uparrow,\downarrow,\uparrow,\downarrow\right)}(\textbf{k},\textbf{Q},\imath\omega_p),&
K_\textbf{q}^{\left(\uparrow,\uparrow,\downarrow,\downarrow\right)}(\textbf{k},\textbf{Q},\imath\omega_p)&0&0\\
K_\textbf{q}^{\left(\uparrow,\downarrow,\uparrow,\uparrow\right)}(\textbf{k},\textbf{Q},\imath\omega_p),&
K_\textbf{q}^{\left(\uparrow,\uparrow,\downarrow,\uparrow\right)}(\textbf{k},\textbf{Q},\imath\omega_p)&0&0\\
K_\textbf{q}^{\left(\downarrow,\downarrow,\uparrow,\downarrow\right)}(\textbf{k},\textbf{Q},\imath\omega_p),&
K_\textbf{q}^{\left(\downarrow,\uparrow,\downarrow,\downarrow\right)}(\textbf{k},\textbf{Q},\imath\omega_p)&0&0
\end{array}%
\right),
\widehat{M}=\left(
\begin{array}{cccc}
0&0&K_\textbf{q}^{\left(\downarrow,\downarrow,\downarrow,\uparrow\right)}(\textbf{k},\textbf{Q},\imath\omega_p),&
K_\textbf{q}^{\left(\downarrow,\uparrow,\uparrow,\uparrow\right)}(\textbf{k},\textbf{Q},\imath\omega_p)\\
0&0&K_\textbf{q}^{\left(\uparrow,\downarrow,\downarrow,\downarrow\right)}(\textbf{k},\textbf{Q},\imath\omega_p),&
K_\textbf{q}^{\left(\uparrow,\uparrow,\uparrow,\downarrow\right)}(\textbf{k},\textbf{Q},\imath\omega_p)\\
0&0&K_\textbf{q}^{\left(\uparrow,\downarrow,\downarrow,\uparrow\right)}(\textbf{k},\textbf{Q},\imath\omega_p),&
K_\textbf{q}^{\left(\uparrow,\uparrow,\uparrow,\uparrow\right)}(\textbf{k},\textbf{Q},\imath\omega_p)\\
0&0&K_\textbf{q}^{\left(\downarrow,\downarrow,\downarrow,\downarrow\right)}(\textbf{k},\textbf{Q},\imath\omega_p),&
K_\textbf{q}^{\left(\downarrow,\uparrow,\uparrow,\downarrow\right)}(\textbf{k},\textbf{Q},\imath\omega_p)
\end{array}%
\right).
\nonumber
\end{equation}
Here, $\omega_{p}=(2\pi/ \beta)p ; p=0, \pm 1, \pm 2,...$ is a Bose
frequency, and we have introduced the two-particle propagator
$K_\textbf{q}^{\left(i,j,k,l\right)}(\textbf{k},\textbf{Q},\imath\omega_p)=
\sum_{\omega_m}G_\textbf{q}^{i,j}(\textbf{k}+\textbf{Q};\imath\omega_p+\imath\omega_m)
G_\textbf{q}^{k,l}(\textbf{k};\imath\omega_m)$, where $
i,j,k,l=\{\uparrow,\downarrow\}$. The condition for existing a
non-trivial solution of the Bethe-Salpeter equations leads to the
following secular determinant
\begin{equation}
Z= \left|
\begin{array}{cccc}
U^{-1}+\left(I_{\gamma,\gamma}-L_{\widetilde{\gamma},\widetilde{\gamma}}\right)&\left(J_{\gamma,l}-K_{m,\widetilde{\gamma}}\right)&
\left(I_{\gamma,\widetilde{\gamma}}+L_{\gamma,\widetilde{\gamma}}\right)&\left(J_{\gamma,m}+K_{l,\widetilde{\gamma}}\right)\\
\left(J_{\gamma,l}-K_{m,\widetilde{\gamma}}\right)&U^{-1}+\left(I_{l,l}-L_{m,m}\right)&
\left(J_{l,\widetilde{\gamma}}+K_{m,\gamma}\right)&\left(I_{l,m}+L_{l,m}\right)\\
\left(I_{\gamma,\widetilde{\gamma}}+L_{\gamma,\widetilde{\gamma}}\right)&\left(J_{l,\widetilde{\gamma}}+K_{m,\gamma}\right)&
-U^{-1}+\left(I_{\widetilde{\gamma},\widetilde{\gamma}}-L_{\gamma,\gamma}\right)&\left(J_{\widetilde{\gamma},m}-K_{\gamma,l}\right)\\
\left(J_{\gamma,m}+K_{l,\widetilde{\gamma}}\right)&\left(I_{l,m}+L_{l,m}\right)&
\left(J_{\widetilde{\gamma},m}-K_{\gamma,l}\right)&U^{-1}+\left(I_{m,m}-L_{l,l}\right)
\end{array}%
\right|,\label{SecDet}\end{equation} where the following symbols are
used:
\begin{eqnarray}&
I_{a,b}=\frac{1}{2N}\sum_\textbf{k}a^\textbf{q}_{\textbf{k},\textbf{Q}}b^\textbf{q}_{\textbf{k},\textbf{Q}}
\left[\frac{1-f\left(\omega_-(\textbf{k},\textbf{q})
\right)-f\left(\omega_+(\textbf{k}+\textbf{Q},\textbf{q})
\right)}{\omega+\Omega_\textbf{q}(\textbf{k},\textbf{Q})-\varepsilon_\textbf{q}(\textbf{k},\textbf{Q})]}
-\frac{1-f\left(\omega_+(\textbf{k},\textbf{q})
\right)-f\left(\omega_-(\textbf{k}+\textbf{Q},\textbf{q})
\right)}{\omega+\Omega_\textbf{q}(\textbf{k},\textbf{Q})+\varepsilon_\textbf{q}(\textbf{k},\textbf{Q})]}\right]
,\nonumber\\&
J_{a,b}=\frac{1}{2N}\sum_\textbf{k}a^\textbf{q}_{\textbf{k},\textbf{Q}}b^\textbf{q}_{\textbf{k},\textbf{Q}}
\left[\frac{1-f\left(\omega_-(\textbf{k},\textbf{q})
\right)-f\left(\omega_+(\textbf{k}+\textbf{Q},\textbf{q})
\right)}{\omega+\Omega_\textbf{q}(\textbf{k},\textbf{Q})-\varepsilon_\textbf{q}(\textbf{k},\textbf{Q})]}
+\frac{1-f\left(\omega_+(\textbf{k},\textbf{q})
\right)-f\left(\omega_-(\textbf{k}+\textbf{Q},\textbf{q})
\right)}{\omega+\Omega_\textbf{q}(\textbf{k},\textbf{Q})+\varepsilon_\textbf{q}(\textbf{k},\textbf{Q})]}\right]
,\nonumber\\&
K_{a,b}=\frac{1}{2N}\sum_\textbf{k}a^\textbf{q}_{\textbf{k},\textbf{Q}}b^\textbf{q}_{\textbf{k},\textbf{Q}}
\left[\frac{f\left(\omega_-(\textbf{k},\textbf{q})
\right)-f\left(\omega_-(\textbf{k}+\textbf{Q},\textbf{q})
\right)}{\omega+\Omega_\textbf{q}(\textbf{k},\textbf{Q})+\epsilon_\textbf{q}(\textbf{k},\textbf{Q})]}
+\frac{f\left(\omega_+(\textbf{k},\textbf{q})
\right)-f\left(\omega_+(\textbf{k}+\textbf{Q},\textbf{q})
\right)}{\omega+\Omega_\textbf{q}(\textbf{k},\textbf{Q})-\epsilon_\textbf{q}(\textbf{k},\textbf{Q})]}\right]
,\nonumber\\&
L_{a,b}=\frac{1}{2N}\sum_\textbf{k}a^\textbf{q}_{\textbf{k},\textbf{Q}}b^\textbf{q}_{\textbf{k},\textbf{Q}}
\left[\frac{f\left(\omega_-(\textbf{k},\textbf{q})
\right)-f\left(\omega_-(\textbf{k}+\textbf{Q},\textbf{q})
\right)}{\omega+\Omega_\textbf{q}(\textbf{k},\textbf{Q})+\epsilon_\textbf{q}(\textbf{k},\textbf{Q})]}
-\frac{f\left(\omega_+(\textbf{k},\textbf{q})
\right)-f\left(\omega_+(\textbf{k}+\textbf{Q},\textbf{q})
\right)}{\omega+\Omega_\textbf{q}(\textbf{k},\textbf{Q})-\epsilon_\textbf{q}(\textbf{k},\textbf{Q})]}\right]
.\nonumber\end{eqnarray}\end{widetext}Here, $\varepsilon _{\boldsymbol{q}}(%
\boldsymbol{k},\boldsymbol{Q})=E_{\boldsymbol{q}}(\boldsymbol{k}+\boldsymbol{%
Q})+E_{\boldsymbol{q}}(\boldsymbol{k})$, $\epsilon _{\boldsymbol{q}}(%
\boldsymbol{k},\boldsymbol{Q})=E_{\boldsymbol{q}}(\boldsymbol{k}+\boldsymbol{%
Q})-E_{\boldsymbol{q}}(\boldsymbol{k})$, $\Omega _{\boldsymbol{q}}(%
\boldsymbol{k},\boldsymbol{Q})=\eta _{\boldsymbol{q}}(\boldsymbol{k})-\eta _{%
\boldsymbol{q}}(\boldsymbol{k}+\boldsymbol{Q})$, and $a$ and $b$ are one of
the following form factors: 
\begin{eqnarray}
\gamma _{\boldsymbol{k},\boldsymbol{Q}}^{\boldsymbol{q}}&=&u_{\boldsymbol{k}%
}^{\boldsymbol{q}}u_{\boldsymbol{k}+\boldsymbol{Q}}^{\boldsymbol{q}}+v_{%
\boldsymbol{k}}^{\boldsymbol{q}}v_{\boldsymbol{k}+\boldsymbol{Q}}^{%
\boldsymbol{q}},  \notag \\
l_{\boldsymbol{k},\boldsymbol{Q}}^{\boldsymbol{q}}&=&u_{\boldsymbol{k}}^{%
\boldsymbol{q}}u_{\boldsymbol{k}+\boldsymbol{Q}}^{\boldsymbol{q}}-v_{%
\boldsymbol{k}}^{\boldsymbol{q}}v_{\boldsymbol{k}+\boldsymbol{Q}}^{%
\boldsymbol{q}},  \notag \\
\widetilde{\gamma }_{\boldsymbol{k},\boldsymbol{Q}}^{\boldsymbol{q}}&=&u_{%
\boldsymbol{k}}^{\boldsymbol{q}}v_{\boldsymbol{k}+\boldsymbol{Q}}^{%
\boldsymbol{q}}-u_{\boldsymbol{k}+\boldsymbol{Q}}^{\boldsymbol{q}}v_{%
\boldsymbol{k}}^{\boldsymbol{q}},  \notag \\
m_{\boldsymbol{k},\boldsymbol{Q}}^{\boldsymbol{q}}&=&u_{\boldsymbol{k}}^{%
\boldsymbol{q}}v_{\boldsymbol{k}+\boldsymbol{Q}}^{\boldsymbol{q}}+u_{%
\boldsymbol{k}+\boldsymbol{Q}}^{\boldsymbol{q}}v_{\boldsymbol{k}}^{%
\boldsymbol{q}}.  \notag
\end{eqnarray}
According to the well-known Goldstone theorem, as $\boldsymbol{Q}\rightarrow
0$, there exists a solution $\omega \rightarrow 0$. In this limit all $J$, $K
$ and $L$ vanish, and the secular equation reduces to the gap equation
written as $0=1+UI_{\gamma =1,\gamma =1}$. 
\begin{figure}[tbh]
\centerline{\epsfig{file=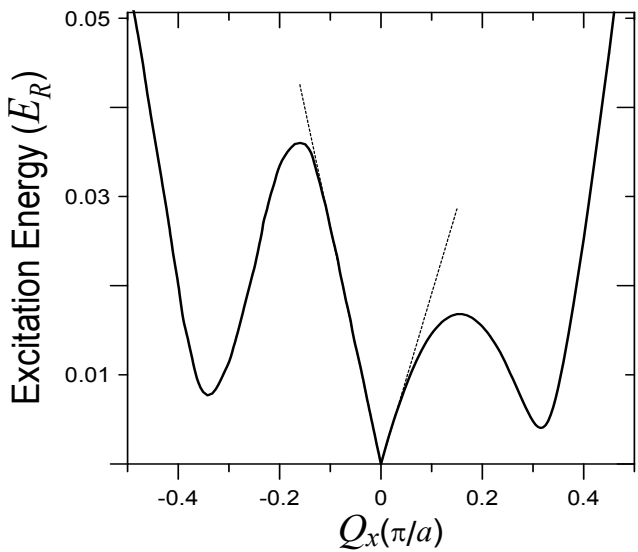,height=3.0in,width=3.5in}}
\caption{Excitation energy for collective modes of a polarized $^{6}$Li gas
in a 3D optical lattice with $\protect\lambda =1030$ nm and total filling
factor $f=0.4685.$ The Hubbard parameters are $J=0.078\ E_{R}$ and the
attractive on-site interaction is $U/J=2.64.$}
\label{fig:CE}
\end{figure}

For $\boldsymbol{Q}\neq 0,$ we use a 3D Gaussian integration in each term in
the secular determinant (\ref{SecDet}) and search for the solution when $%
Z=0. $ Without loss of generality, we fix the collective excitation momentum 
$\boldsymbol{Q}$ in the $x-$direction, $(Q_{x},0,0)$.  For small values of $Q_{x}$ the excitation energy is the
linear, low-energy (Goldstone) mode in the FFLO state corresponding to the
fluctuations of the order parameter phase, but since the FFLO state breaks
both gauge and translational symmetry there are two distinct modes as shown
in Fig. \ref{fig:CE}. In this case, the polarization is $%
P=0.093883$; the filling-fraction parameters are $%
f_{\uparrow }=0.256248$ and $f_{\downarrow }=0.212263$, and $U/J=2.64$ at a
temperature $k_{B}T/E_{R}=10^{-4}.$ The two distinct sound velocities in the
long wavelength limit are 8.56 mm/s and 6.14 mm/s as shown for the negative
and positive wavenumbers, respectively. The results from our numerical
solutions of the BS equation also show that the Goldstone modes have
rotonlike minima, $\omega_{r}=0.0077E_{R}$ and $\omega_{r}=0.004E_{R}$.

In Fig.  \ref{fig:CE}, the rotonlike structure is clearly seen and the minimum requirements on the
flow velocities to be able to slow down (obtained from the two roton slopes)
are $v_{1}=0.725$ mm/s and $v_{2}=0.41$ mm/s, respectively. The asymmetry of
the sound mode and the roton minima originates from the fact that the
population imbalance is achieved when either $\omega _{+}(\boldsymbol{k}+%
\boldsymbol{Q},q_{x})$ or $\omega _{-}(\boldsymbol{k}+\boldsymbol{Q},q_{x})$
is negative in some regions of momentum space, but the regions are different
for positive and negative $Q_{x}$. The answer of the question how this
asymmetry is related to $f_{\uparrow },f_{\downarrow }$ and $U/J$ requires
analytical expressions for the two regions which is beyond the goals of the
present work.

\section{Conclusions}

In this paper we have presented the phase diagram and the collective
excitations of an imbalanced system of $^{6}$Li atoms loaded in a cubic
optical lattice. Upon minimization of the free energy, the stability regions
of BCS, Sarma, FFLO and BCS-normal mixed-state phases were obtained. We also showed
that the FFLO phase can be quite large compared to both, the Sarma
and the phase separation regions when the hopping strength in the single-band
Hubbard model is increased and the on-site attraction is decreased. The
effects of dimensionality were also analyzed by contrasting the phases of a
system loaded in a 3D optical lattice with an identical, 2D system where we
showed that the lower dimensionality gas can sustain larger polarizations in
the FFLO phase.

We also derived a Bethe-Salpeter equation for the attractive Hubbard
Hamiltonian based on the generalized random phase approximation to calculate the
collective mode spectrum of the Fermi gas in a deep optical lattice. Using a
contact interaction, an algebraic equation for the BS amplitudes was
obtained. The solution for the excitation spectrum of collective modes was
derived by calculating the roots of the corresponding secular $4\times 4$
determinant. For $\boldsymbol{Q}\rightarrow 0$ we obtained two distinct
Goldstone modes and  their respective sound velocities. For shorter
wavelengths, we showed that the Goldstone modes have an asymmetric rotonlike
spectrum. The critical flow velocities in this region were calculated to
show that superfluidity can survive in a polarized fermion gas in two- and
in three-dimensional optical lattices.

\bigskip 

This work was partially supported by UNAM-DGAPA grants IN-105011 \&
IN-111613, and Conacyt 104917.

\end{document}